\documentclass[twocolumn,floats,epsfig,pdflatex]{revtex4}
\usepackage{graphics,epsfig,graphicx}
\usepackage{color}
\usepackage{makeidx}
\usepackage{latexsym}

\def\bra#1{\mathinner{\langle{#1}|}}
\def\ket#1{\mathinner{|{#1}\rangle}}

\begin{document}

\title{Berry Phase Generation and Measurement in a Single Trapped Ion}

\author{D. De Munshi,$^{1}$ M. Mukherjee,$^{1}$\footnote{E-mail id: phymukhe@nus.edu.sg (corresponding author)} B. Dutta Roy\footnote{E-mail id: bnykdr@gmail.com} }

\affiliation{$^{1}$Center for Quantum Technologies, National
University of Singapore, Singapore - 117543}

\date{\today}

\begin{center}
\begin{abstract}

In this work, we propose a new design of an ion trap which can
enable us to generate state specific Berry phase in a single trapped
ion. Such a design will enable us to study the physics at the
boundary of abelian and non-abelian symmetries and can also have
significant impact in quantum computation.

\end{abstract}
\end{center}

\maketitle

\section{Introduction}

Single trapped ion provides the cleanest quantum system available in
nature. In the past decade or more, trapped ions has been used in a
variety of scenarios, from developing atomic clocks in the optical
domain \cite{diddams2, ybclock, srclock} to implementing protocols
required for quantum computation \cite{Monr95}. It also provides a
test bed for measurement of various fundamental properties of
quantum mechanics and quantum field theories, such as parity
non-conservation and measurements on limits of electron electric
dipole moment\cite{Nov93,Man10,Lea11} as well as
time variation of fundamental constants of nature\cite{Vvflam99}. \\

Geometric phases in general and Berry phase in particular has been
studied for a variety of physical systems and conditions
\cite{Hors07, Wilc84, Avro87, Wang99}. The theoretical treatment of
geometric phases has been generalized for composite systems having
internal structure particularly in the light of quantum optics
\cite{Meye09}. This has practical importance from the view point of
systemic effects associated with rotating electric and magnetic
fields in precision experiments \cite{Vuth09}. However, higher order
multipole effects has not been considered except for a few cases like
the Nuclear Quadrupole Resonances in a rotating frame \cite{Tyc87}
where frequency shift in the NQR spectrum due to Berry phase has
been observed.
In this work, we put forward a proposal for the measurement of Berry
phases and Berry phase generated energy shifts using single trapped
ion.
The ion trap experiment will not only provide observable energy shift but also
a clean and controlled measurement to understand the geometric phase
in Abelian and non-Abelian cases. To the best of our knowledge, this
has so far not been explored either theoretically or experimentally.
In a way, this proposal allows one to simulate a situation when a
quantum system changes from one symmetry to another symmetry. In
either regime the Berry phases may be used to implement various
noise tolerant quantum gates, which form the heart of
quantum computation, without the complexity of using lasers.\\

In this work, we show that we can generate Berry phase and its
associated energy shifts using modified ion trap geometry. We also
present a convenient technique for calculation of Berry phases in
situations involving electric quadrupole moment interacting with a
time dependent electric field gradient.

\section{Berry Phase and Phase Dependent Energy Shift}

\subsection{Introduction to Berry Phase}

Berry phase is a phase acquired by the eigenstates of a Hamiltonian,
which is changing with time implicitly in an adiabatic fashion
\cite{Ber84} in a parameter space so that the system remains in the
same eigenstate during the entire evolution, \textit{i.e.} the time
dependence does not lead to a transition in the system. For
adiabatic condition to hold, time scale of the change must be less
than any other relevant time scale of the system.  The Schr\"odinger
equation for the adiabatic evolution of a system can be written as

\begin{equation}
H(\overrightarrow{R}(t))
\psi_{n}(\overrightarrow{R}(t))=E_{n}(\overrightarrow{R}(t))
\psi_{n} (\overrightarrow{R}(t)),
\end{equation}

where the time dependence comes through the parameter $R(t)$. Under
such time evolution, the total wavefunction is given by

\begin{equation}
\ket{\psi_n(\overrightarrow{R}(t))}=e^{\frac{i}{\hbar}\int^{t}_{0}
E_{n}(\overrightarrow{R}(t)) dt} ~e^{i
\gamma_{n}(t)}~\ket{n(\overrightarrow{R}(t))}, \label{fullwf}
\end{equation}

where $\gamma_{n}(t)$ is the Berry phase acquired in time $t$ by the $\ket{n(\overrightarrow{R}(t))}$ eigenstate and
$E_{n}(\overrightarrow{R}(t))$ is the corresponding eigenvalue. For
a periodic time evolution of the Hamiltonian (which is the most
practical of experimentally viable scenarios), the Berry phase
acquired by the $n$th eigenstate is given by

\begin{equation}
\gamma_0^n= i \oint \bra{\psi(\overrightarrow{R})}
\nabla_{\overrightarrow{R}} \ket{\psi(\overrightarrow{R})} \cdot
d\overrightarrow{R},\label{bqeq}
\end{equation}

where the integration is carried over the path traversed in the
parameter space.\\

\subsection{Phase Dependent Energy Shift}
The phase of the wavefunction in Eq. \ref{fullwf} in the linear
approximation of variation of Berry Phase with time due to adiabatic
change can be re-written as

\begin{equation}
\frac{i}{\hbar} \int_{0}^{t}(E_{n}(\overrightarrow{R}(t)) +
\frac{\hbar \gamma_{0}^{n}}{T}) dt, \label{modphase}
\end{equation}

where $ \gamma_{0}^{n}$ is the Berry Phase acquired in one cycle with
a time period $T$.\\

As can be seen from Eq. \ref{modphase}, the eigenvalue of the system
gets modified by $\frac{\hbar \gamma_{0}^{n}}{T}$. Thus in presence
of a periodically evolving time dependent Hamiltonian, adiabatic in nature,
the energy eigenvalue of the Hamiltonian gets modified depending on
the acquired phase and the time period of the cyclical evolution.

\section{Proposed Experimental System}

In this work we propose a concrete experiment where we can observe
such phase dependent energy shifts using a single trapped ion. The
trap is modified to impose a cyclic Hamiltonian on the ion and we
can observe the resultant shift in the electronic levels of the
trapped ion. The interaction by which we propose to impose such a
time dependent Hamiltonian is the interaction of the quadrupole
moment of some chosen electronic levels with a time dependent
electric field gradient,provided by a modified trap geometry. To the
best of our knowledge, the only experimental observation of similar
splitting and shift has been by R. Tycko who used a single crystal
of KClO$_3$, where the crystal field gradient interacted with the
nuclear quadrupole moment and the time dependence has been
incorporated by mechanical rotation of the crystal. The proposed
experiment using a single trapped ion however is fundamentally
different in two ways. First, the electronic state of the ion allows
the
manipulation of the Hamiltonian by light field. Second, the presence or absence of static magnetic field allows the system to change between Abelian and non-Abelian geometry. This transition from one regime to another has so far not been studied either experimentally or theoretically.\\

\subsection{Calculation of Berry Phase}

For the generation of a Berry phase, we need a time dependent field
gradient and a system with a quadrupole moment. In this case the
time dependent field gradient is provided by the modified trap
design.\\

The quadrupole moment can be defined as

\begin{equation}
Q_{ij}=c (\frac{1}{2}(S_i S_j - S_j S_i) -\frac{1}{3}
\overrightarrow{S}^2)
\end{equation}

where $S_{k}$ corresponds to the k$^{th}$ component of the spin and $c$ is a numeric constant. The Hamiltonian of the interaction is given as

\begin{equation}
H=\frac{1}{6} Q_{ij}\frac{\partial E_{i}}{\partial x_{j}},
\end{equation}

where $\frac{\partial E_{i}}{\partial x_{j}}$ is the $ij$the
component of the electric field gradient tensor. Now, we first move
into the frame of the field gradient and define the $z^{'}$-axis
along the field gradient tensor. In that case, the Hamiltonian
becomes

\begin{equation}
H=\alpha (S_{z^{'}}^2-\frac{1}{3} \overrightarrow{S}^2),
\end{equation}

where the pre-factor $\alpha$ contains the value of the electric
field gradient and the quadrupole moment of the state under study.
Assuming a spin $\frac{3}{2}$ state, which is relevant, as the
experimental system we are interested in is the $5d_{\frac{3}{2}}$
state of a trapped Ba+ ion, the eigenvalues for this Hamiltonian are
$\alpha\hbar^2$ and $-\alpha \hbar^2$ with each of them being doubly
degenerate due to Kramers degeneracy. $|\frac{3}{2}>$ and
$|-\frac{3}{2}>$ are the first set of doubly degenerate eigenstates
and $|\frac{1}{2}>$ and $|-\frac{1}{2}>$ consists of the second set.
The value of alpha will depend on the value of the quadrupole moment
of the $5d_{\frac{3}{2}}$ of Ba$^{+}$ ion \cite{Ita06}, as well as
the magnitude of the components of electric field gradient.

Now to transform back to the laboratory frame, we apply the Wigner D-matrices. Transformation to the laboratory frame consists of a rotation of $-\phi$ along the z-axis and then $-\theta$ along the
rotated y axis and the third rotation being a null rotation. The time dependence comes from $\phi=\omega t$. Applying the Wigner matrices and using Eq. \ref{bqeq}, we obtain the following Berry phases
for the eigenstates of the Hamiltonian

\begin{figure}
\vspace{10mm} \hspace{.5cm}
  \includegraphics[scale=.5]{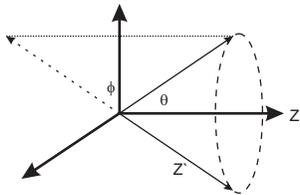}\\
  \caption{Principle axis of field gradient relative to laboratory axis. $\phi=\omega t$ is the time dependent parameter leading to rotation of the principle axis w.r.t. lab z-axis}
  \label{grad_axis}
\end{figure}

\begin{eqnarray}
\gamma_{4}=-3 \pi( \cos\theta -1) \nonumber\\
\gamma_{3}=-\pi((4-3 \cos^2 \theta)^{\frac{1}{2}}-1) \nonumber\\
\gamma_{2}=\pi((4-3 \cos^2 \theta)^{\frac{1}{2}}-1) \nonumber\\
\gamma_{1}=3 \pi( \cos\theta -1),
\end{eqnarray}

where $4$, $3$, $2$ and $1$ signify the states $\ket{3/2}$, $\ket{1/2}$, $\ket{-1/2}$ and $\ket{-3/2}$ respectively.\\
The constants of integration are so chosen as to ensure zero phase
for $\theta=0$. Thus incorporating the phase dependent energy
shifts, we obtain the energies of the eigenstates as

\begin{eqnarray}
E_{4}=\alpha \hbar^2 - \frac{\hbar 3 \pi( \cos\theta -1)}{T} \nonumber \\
E_{3}=-\alpha \hbar^2 - \frac{\hbar \pi((4-3 \cos^2 \theta)^{\frac{1}{2}}-1)}{T} \nonumber\\
E_{2}=-\alpha \hbar^2 + \frac{\hbar \pi((4-3 \cos^2 \theta)^{\frac{1}{2}}-1)}{T} \nonumber\\
E_{1}=\alpha \hbar^2 + \frac{\hbar 3 \pi( \cos\theta -1)}{T}.
\label{eqEng}
\end{eqnarray}

Thus the eigenvectors split depending on the Berry phase of each of the states as well
as the frequency of rotation of the Hamiltonian. The Berry phase itself is independent of any external field value but only depends on the geometry of the rotation axis with respect to the quantization axis. This is not surprising since this phase is a purely geometric in nature and is potentially applicable to perform fault tolerant quantum information processing. \\

\subsection{Application of Rotating Field Gradient}

\begin{figure}
\vspace{10mm} \hspace{.cm}
  \includegraphics[scale=.5]{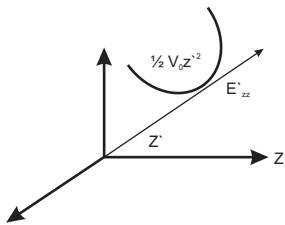}\\
  \caption{The tilted potential $\frac{1}{2}V_{0}z'^{2}$ and resultant field gradient $E'_{zz}$ }
  \label{potgrad}
\end{figure}

The proposal is based on the fact that a potential tilted with
respect to the trap axis ($\frac{1}{2}V_{0}z'^{2}$) can give rise to
a field gradient whose principal axes are tilted with respect to the
trap symmetry axes as shown in Fig \ref{potgrad}.However it has ben
shown by our calculation and also in Ref \cite{Ita06} that the main
contributing component is the $\frac{\partial E_{z}}{\partial z}$
component. It can be proved by taking the potential $V= \frac{1}{4}
V_{0} x^{2} + \frac{1}{4} V_{0} y^{2} - \frac{1}{2} V_{0} z^{2}$ and
calculating the Hamiltonian of quadrupole interaction. Hence for
simplicity, we consider only the $\frac{\partial E_{z}}{\partial z}$
component of the field gradient matrix to be non-zero with the other
components to be zero. The proof of the fact that a tilted potential
can give rise to a tilted field gradient will be obvious by
calculating the field gradient matrix from a potential $\frac{1}{2}
V_{0} z^{' 2}$ and comparing it with a matrix obtained by rotating
the basis of a gradient matrix with only $z^{'}z^{'}$ component and
representing it in the trap basis, i.e. $R^{T} E^{'}_{z^{'}} R$ with
Euler angles $\phi$,$\theta$ and $0$.

\begin{figure}
\vspace{10mm} \hspace{.cm}
  \includegraphics[scale=.25]{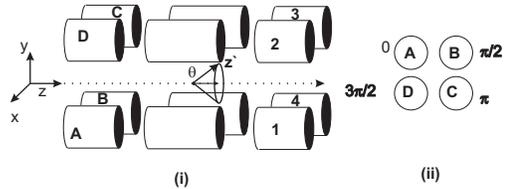}\\
  \caption{(i) Trap design and the resultant field gradient axis. (ii) Cross-sectional view and phase of applied RF. }
  \label{trapgeo}
\end{figure}

Thus the design goal is to have electrode geometry which can produce a parabolic
potential rotating about the linear trap axis. It can be established using a four rod structure for the end caps in a linear paul trap and applying RF voltages on diagonally opposite end cap rods across the body. As is demonstrated in figure \ref{trapgeo}, by connecting body diagonally opposite end cap rods (e.g. $A$ and $3$) and applying suitably phase shifted RF voltage on each of the four pair of end cap rods ($\frac{\pi}{2}$), it is possible to rotate a parabolic potential about the trap axis \cite{Hua97}, leading to rotation of the field gradient about the axis of symmetry.\\

\begin{figure}
\vspace{0mm} \hspace{.cm}
  \includegraphics[scale=.2]{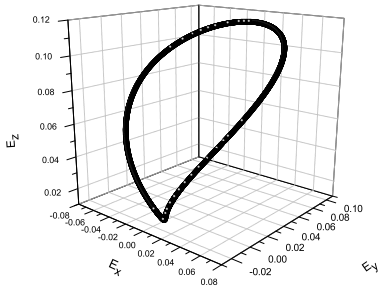}\\
  \caption{Phase diagram of time dependent electric field components according to theoretical predictions. }
  \label{sim_elec}
\end{figure}

\begin{figure}
\vspace{0mm} \hspace{.cm}
  \includegraphics[scale=.2]{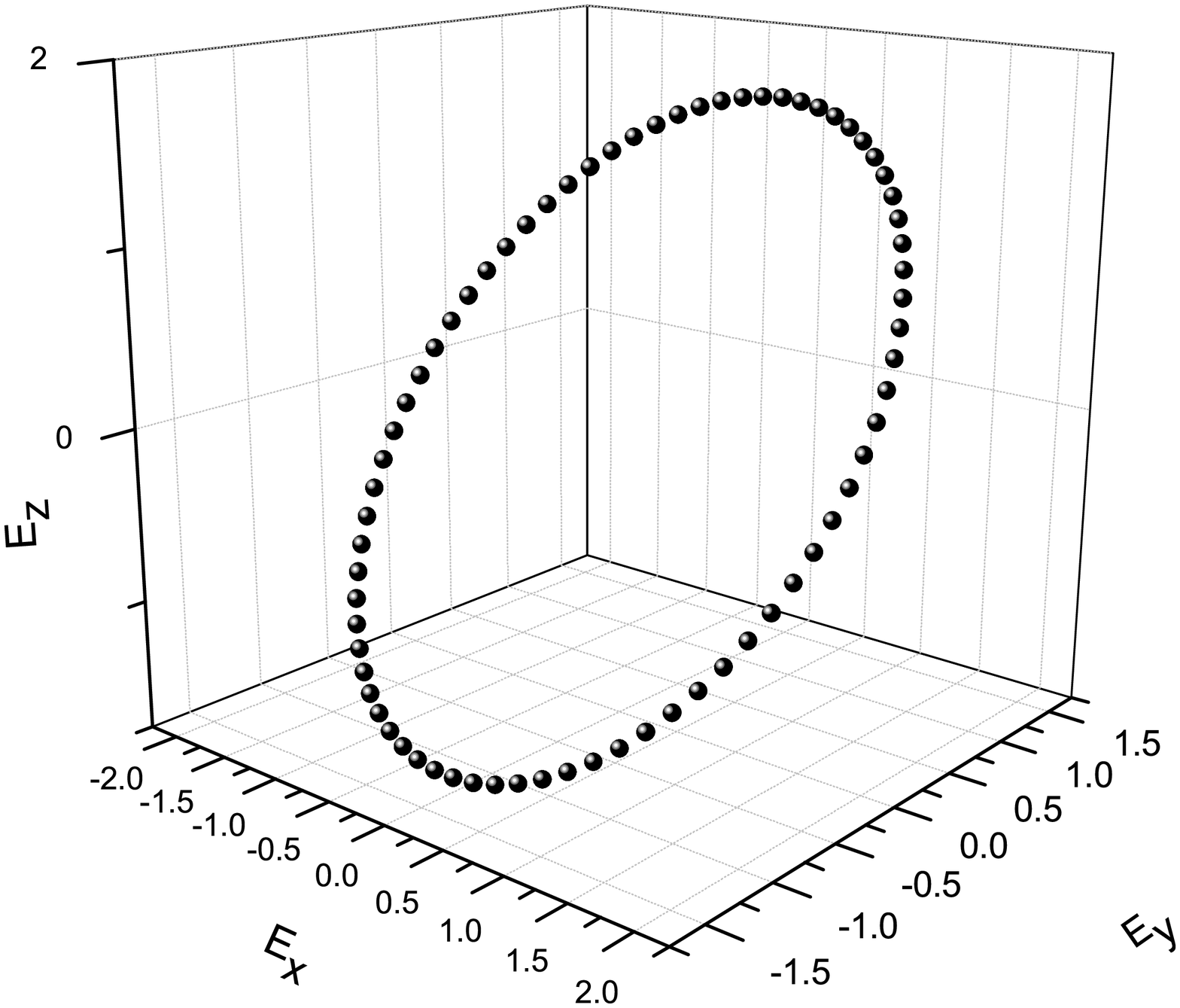}\\
  \caption{Phase diagram of time dependent electric field components from Simion.}
  \label{trap_elec}
\end{figure}

Here we will show that such a distorted trap indeed leads to a tilted potential. We simulate the trap potential, given its geometry and the rotating time dependent potential, using Simion 7.0
and plot the three components of the electric field for a position, off axis from the trap center (exactly at the trap center the field is zero). Also we plot the electric field as obtained theoretically by taking the gradient of the potential $\frac{1}{2}V_{0}z'^{2}$. The electric field components, $E_{x}$,$E_{y}$ and $E_{z}$ are plotted as functions of each other in a three dimensional graph as a
phase diagram and the two situations are compared, Fig. \ref{sim_elec} and Fig. \ref{trap_elec}.\\

As can be seen there are certain differences between the two figures. They arise out of the fact that even though $\frac{1}{2}V_{0}z'^{2}$ assumes a $x'$ and $y'$ symmetry, the actual trap does not have so and hence it can lead to the difference that is observed. The value of $\theta$ has been taken as $40.7^{o}$ for generating the first graph. This value corresponds to the angle a straight line, connecting the middle of two diagonally opposite rods, makes with the trap axis. Hence we can assume that the trap creates a sort of a distorted parabola when the above
mentioned potential is applied to the end caps.\\

\begin{figure}
\vspace{0mm} \hspace{.cm}
  \includegraphics[scale=.2]{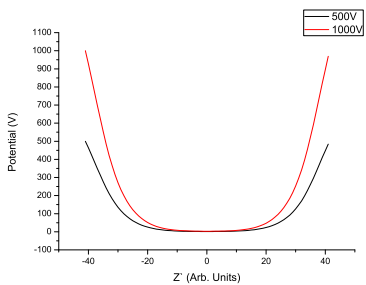}\\
  \caption{Potential along $z'$ for end cap voltage of 500 and 1000.}
  \label{pot500}
\end{figure}

To obtain the true form of the potential in the case of the actual trap geometry, the potential along the diagonal is plotted for two values of the endcap voltages $500$V and $1000$V  respectively (Fig. \ref{pot500}). A fourth order polynomial has been fitted in both the cases.\\

The fitting parameters show that as the voltages of the end caps are
increased, the parabolic nature of the potentials also increase.
However, the berry phase and the resultant energy difference is
independent of the magnitude of the field gradient and hence it should not affect the splitting of the levels.\\
The frequency of rotation however has to be very slow. This is
because, to maintain the adiabatic condition, the precession
frequency has be much smaller than the level splitting caused by the
Hamiltonian itself. As an example, for Ca ion, the splitting is of
the order of $150$ Hz for an electric field gradient of about $50$V/mm$^2$\cite{Roo06}. Since the quadrupole moment of
Ba$^{+}$ is almost double that of Ca$^{+}$ \cite{Ita06}, hence we
can assume that given similar gradient magnitudes, the splitting
will be of the order of $300$ Hz and the rotation frequency should
be much less than that. For such low frequencies, it will not affect
the motion of ions in the trap as the relevant trap frequencies are
of the order of MHz.

\section{Applications}

From the point of view of Abelian and Non-abelian physics, this
system can provide insights into transition from abelian to
non-abelian situations. For example, when abelian situation is
there, that is the states are degenerate, the phase acquired by the
$|\frac{1}{2}>$ and $|-\frac{1}{2}>$ sub-states of the
$5d_\frac{3}{2}$ state are $\pm \pi((4-3 \cos^2
\theta)^{\frac{1}{2}}-1)$, whereas if they non-degenerate, for
example, by applying small symmetry breaking magnetic field, the
phases should become $\pm \pi (\cos \theta -1)$. This is because of
the prescription by Wilczek and Zee \cite{Wilc84} where the off
diagonal terms of the phase matrix are considered only if the states
are non-degenerate. If there is a physical manifestation of it, then
for the abelian and non-abelian scenario, the phase dependent energy
shifts will be different.

\section{Conclusion}

In this article we have shown that an ion trap with a modified
geometry can be used to generate observable splitting which are purely
geometric in nature, on meta-stable D-states of ions. The advantage
of using the electric quadrupole moment is that it will lead to
selective splitting, occurring only for states which have quadrupole
moments. The geometric nature of the acquired phases as well as the
state specificity of the phases make this system quite attractive
for quantum information processing. This type of shifts can be considered as possible systematic in precision experiments dealing with rotating fields and their gradients.
As a conclusion the proposed experiment is the first direct attempt
to test quantum Physics in an interface of two symmetries namely,
Abelian and non-Abelian.

\section{Acknowledgements}
The authors would like to thank Sir M.V. Berry for many useful
discussions regarding Berry phase. DDM and MM would also like to
thank DST, India for financial support.

\end{document}